\documentstyle[aaspp4,flushrt]{article}
\begin{document}
\title{Hydrodynamical Simulations of Corotating Interaction Regions
and \\ Discrete Absorption Components in Rotating O-Star Winds}
\author{Steven R. Cranmer and Stanley P. Owocki}
\affil{Bartol Research Institute,
University of Delaware,
Newark, DE  19716  \\
I:  cranmer@bartol.udel.edu, owocki@bartol.udel.edu}

\begin{abstract}
We present two-dimensional hydrodynamical simulations of corotating stream
structure in the wind from a rotating O~star, together with
resulting synthetic line profiles showing discrete absorption
components (DACs).
An azimuthal variation is induced by a local increase or decrease
in the radiative driving force, as would arise from a bright or
dark ``star spot'' in the equatorial plane.
Since much of the emergent wind structure seems independent of the
exact method of perturbation, we expect similar morphology in
winds perturbed by localized magnetic fields or nonradial pulsations,
as well as by either rotationally-modulated structure or transient
mass ejections.

We find that bright spots with enhanced driving
generate high-density, low-speed streams, while
dark spots generate low-density, high-speed streams.
Corotating interaction regions (CIRs) form where fast material
collides with slow material -- e.g. at the leading (trailing) edge of
a stream from a dark (bright) spot, often steepening into shocks.
The unperturbed supersonic wind obliquely impacts the high-density CIR
and sends back a nonlinear signal which takes the form of a
sharp propagating discontinuity
(``kink'' or ``plateau'') in the radial velocity gradient.
These features travel inward in the co-moving frame at the radiative-acoustic
characteristic speed, and thus
slowly outward in the star's frame.
We find that these slow kinks, rather than the CIRs themselves,
are more likely to result in high-opacity DACs in the absorption
troughs of unsaturated P~Cygni line profiles.
Because the hydrodynamic structure settles to a steady state in a
frame corotating with the star, the more tightly-spiraled kinks
sweep by an observer on a longer time scale than material moving
with the wind itself.
This is in general accord with observations showing slow apparent
accelerations for DACs.
\end{abstract}


\section{Introduction}

The radiatively-driven winds of early-type (O, B, Wolf-Rayet) stars
are observed to vary on time scales ranging from hours to years.
In addition, as in the highly complex solar wind, the mass outflows
from hot stars are presumably not spherically symmetric.
There are many physical mechanisms that can lead to wind structure
and variability, and it is useful to distinguish between
(1) {\em small-scale} stochastic fluctuations, intrinsic to the
wind itself, and
(2) {\em large-scale} quasi-regular variability, induced by changes
in the underlying star.
In the former category is the the shocked structure arising from
the strong instability of the line-driving mechanism
(Rybicki \markcite{R87}1987; Owocki \markcite{O92}1992),
which may explain black troughs in saturated UV P~Cygni lines in
OB~stars (Lucy \markcite{L82}1982; Puls, Owocki, \& Fullerton
\markcite{POF}1993, hereafter POF),
shock-heated X-ray emission (Cooper \& Owocki \markcite{CO94}1994),
and moving ``bumps'' in Wolf-Rayet optical emission lines
(Robert \markcite{R94}1994).
On the other hand, the larger-scale structure could be attributed to
the dynamical effects of rotation,
magnetic fields, or nonradial pulsations, which
may produce the recurring discrete absorption components (DACs)
and blue-edge variability observed in ultraviolet P~Cygni lines.
There has been much progress in using radiation hydrodynamics to model
the small-scale intrinsic wind instability, but
considerably less attention has been given to the problem of how
line-driven winds respond to larger star-induced
variations.
This paper reports on initial results from radiation hydrodynamics
models of winds affected by such laterally coherent and
rotationally modulated perturbations.

Large-scale wind structure in hot stars is inferred most directly
from time variability in
the blueshifted absorption troughs of UV P~Cygni profiles.
The most conspicuous variations are the DACs, which
appear as narrow and localized optical depth enhancements
in unsaturated lines,
in some stars even dominating the ``mean wind'' absorption.
DACs are present in a majority of
O-star (Howarth \& Prinja \markcite{HP89}1989) and
Be-star (Grady, Bjorkman, \& Snow \markcite{Ge87}1987) winds,
and are typically seen to accelerate to the blue wing of the profile
over a few days, becoming narrower as they approach an asymptotic
velocity.
Prinja \markcite{P88}(1988) and Henrichs, Kaper, \& Zwarthoed
\markcite{HKZ}(1988) found an
apparent correlation between both the recurrence and acceleration
time scales of DACs
(typically of the same order as each other)
and the projected rotational velocity of the star,
$V_{\mbox{\footnotesize eq}} \sin i$.
Corresponding and often temporally-correlated variability is seen
in the blue edge fluctuations of
saturated UV P~Cygni lines,
in the low-velocity variability of subordinate-level P~Cygni lines,
and in optical lines such as $\mbox{H}\alpha$ and
\ion{He}{2} $\lambda 4686$,
suggesting a single dynamical phenomenon reaching down to very near
the photosphere
(Henrichs, Kaper, \& Nichols \markcite{HKN}1994).

Attempts to model DACs have been progressively constrained
by better observations.
By studying lines of different ionization species, Lamers, Gathier,
\& Snow \markcite{Le82}(1982)
ruled out the early supposition that DACs might be caused by ionization
gradients in an otherwise spherically symmetric and time steady wind.
The episodic ejection of spherical ``shells'' of increased mass loss
was an often-invoked model for a time, but the lack of both
{\em emission} variability in UV P~Cygni lines
(Prinja \& Howarth \markcite{PH88}1988) and significant infrared
variability (Howarth \markcite{H92}1992)
seems to rule out a spherically-symmetric disturbance.
On the other hand, to produce the observed strong absorption dips,
the structure must be large enough to cover a substantial fraction
of the stellar disk.
This seems to rule out the small-scale wind instability as the source
of most DAC clumpiness,
since global averaging would weaken the observable signature
(Owocki \markcite{O94}1994).
Also, Rybicki, Owocki, \& Castor \markcite{Re90}(1990) showed that
small-scale, lateral velocity perturbations should be
strongly damped, and so should not disrupt the horizontal scale
size set by base variations.
Altogether, these constraints suggest that DACs originate from
moderate size wind structures,
e.g.,  spatially-localized clouds, streams, or ``blobs.''

Of particular interest is the apparent acceleration rate of DACs.
When compared to the acceleration of the mean wind inferred from
line-driven flow theory and detailed profile fitting,
some (typically weaker) DACs seem to be passively carried along the same
velocity law (see, e.g., Kaper \markcite{K93}1993).
But most strong DACs accelerate much more {\em slowly} (Prinja
\markcite{P94}1994),
suggesting they may not represent a single mass-conserving feature,
but rather might arise from a slowly evolving {\em pattern} or perturbation
through which wind material flows.
The enhanced optical depth could result from either a higher density
or a lower wind velocity gradient (a ``plateau''), or by a combination
of the two (Fullerton \& Owocki \markcite{FO92}1992;
Owocki, Fullerton, \& Puls \markcite{Oe94}1994),
as is found in the dynamical models below (\S~3).

Mullan (\markcite{M84a}1984a, \markcite{M84b}b; \markcite{M86}1986)
proposed that DACs and related phenomena
could arise from ``corotating interaction regions'' (CIRs) analogous
to those commonly observed {\em in situ} in the solar wind.
In the solar corona, regions of open magnetic field cause the
flow from coronal holes to be accelerated faster than the mean
ecliptic-plane wind, resulting in colliding fast and slow streams strung
into spiral CIR patterns by rotation (Hundhausen
\markcite{H72}1972; Zirker \markcite{Z77}1977).
These nonlinear interacting streams eventually steepen into
oblique corotating shocks, through which the wind flows nearly radially.
Because hot stars do not have the strong surface convection
and coronae known to exist in the sun, the ``seed'' mechanism for
large-scale azimuthal perturbations may be quite different.

In this paper we do not adhere to any particular model
for these photospheric variations,
but several plausible scenarios have been proposed.
Underhill \& Fahey \markcite{UF84} (1984) and Henrichs et
al.\  \markcite{HKN}(1994) suggested
that small patches of enhanced
magnetic field could exist undetected on early-type stars and
produce corotating wind structure.
Also, nonradial pulsations have been observed in many OB stars, and
have been shown to be able to induce
localized increased mass loss and outward
angular momentum transfer (Castor \markcite{C86}1986; Willson
\markcite{W86}1986; Ando \markcite{A91}1991).
Circumstellar disks exhibit many natural large-scale instabilities,
e.g.\  Okazaki's \markcite{O91}(1991) global one-armed
normal modes, which may be correlated with DAC variability in
Be stars (Telting \& Kaper \markcite{TK94}1994).
Dowling \& Spiegel \markcite{DS90}(1990) discuss the possible existence of
Jupiter-like zonal bands and vortices in the atmospheres of hot
stars, and give order-of-magnitude estimates of the flux
enhancement over a ``Great Red Spot'' type of shear pattern.

Several qualitative attempts have been made to apply the CIR
picture to observations of time variability in early-type stellar
winds, but all have been {\em kinematic} in nature.
Prinja \& Howarth \markcite{PH88}(1988)
fit slowly-accelerating spiral streamlines
to DACs in time series spectra from the O7.5 giant 68~Cyg, and showed
that the narrowing of the absorption feature as it accelerates can
be explained roughly by the decrease in the line-of-sight velocity
gradient of the CIR.
Harmanec \markcite{H91}(1991) extended this analysis and discussed possible
observational signatures of CIRs in other classes of early-type stars.
Rotationally-modulated gas streams or ``spokes'' have been proposed
in models of Be star circumstellar material
(see, e.g., \v{S}tefl et al.\  \markcite{Se95}1995)
and in the time variability of Herbig
Ae star spectra (Catala et al.\  \markcite{Ce91}1991).
Of course, the physics of circumstellar streams in Be and Herbig Ae stars
will most likely be very different from that of corresponding
structures around O stars, and we will focus mainly on the latter.

The principal goal of this paper is to model the
{\em dynamical} effect of radiative
driving on the formation of CIRs in a hot star wind,
and to compute synthetic observational diagnostics to see if, e.g.,
slow DAC-like signatures can be theoretically produced.
Our computational approach is to apply a reasonable parameterization
for a localized ``star spot'' perturbation in the radiation force near
the stellar surface, and allow the wind to respond consistently.
Although the actual photospheric structure perturbing the base of
the wind is likely to be different,
we suspect the characteristic response
of a radiatively-driven medium will be insensitive to the details of
this physical mechanism,
and mainly depend on base changes in the fluid velocity and density.

The remainder of this paper is organized as follows.
We first (\S~2) describe our radiation hydrodynamics code and the
details of the induced azimuthally-dependent force enhancement.
Next (\S~3) we present results for a series of O~star models with
varying rotation rates and wind parameters, and discuss the emergent
corotating structure.
We then (\S~4) compute synthetic UV P~Cygni line profile time series
for the various models in our parameter study.
Finally, a discussion and conclusion section
(\S~5) summarizes our results and
outlines directions for future work.

\section{Numerical Radiation Hydrodynamics}

\subsection{Equations of Hydrodynamics}

The problem of structure formation in a stellar wind is
in general arbitrarily three-dimensional.
Rotation imposes a latitudinal dependence on both the photosphere
and the wind, and large-scale variations at the stellar surface
can impose an arbitrary latitudinal or azimuthal
dependence on the wind.
However, to study how a line-driven wind reponds to rotationally
induced variations within a more tractable, {\em two-dimensional} model,
we confine the simulations here to the {\em equatorial plane,}
where rotation has the strongest impact, and where the flow can be
constrained to a surface of constant colatitude, $\theta = \pi /2$.
(See Pizzo \markcite{P82}1982 for discussion of similar approximations
in modeling the solar wind.)
This assumption naturally suppresses the centrifugal wind compression
effect of Bjorkman \& Cassinelli \markcite{BC93}(1993), which is
only of minor importance for O stars.
The induced ``spot'' variations thus
only have an azimuthal extent, and assumptions about
latitudinal structure are only required when computing
observational diagnostics (\S~4), not the inherent dynamics.

We use a time-dependent numerical hydrodynamics code to evolve a
model of a radiatively-driven wind from a rotating star toward an
equilibrium corotating steady state.
The code, VH-1, was developed by J.\  M.\  Blondin and colleagues
at the University of Virginia,
and uses the piecewise parabolic method (PPM) algorithm developed by
Collela \& Woodward \markcite{CW84}(1984).
VH-1 solves the Lagrangian forms of the equations of hydrodynamics
in the fluid rest frame, and remaps conserved quantities onto an
Eulerian grid at each time step.
The equations to be numerically integrated, written in Eulerian form
using spherical polar coordinates, include the conservation of mass,
\begin{equation}
  \frac{\partial \rho}{\partial t} +
  \frac{1}{r^2} \frac{\partial}{\partial r}
  \left( \rho v_{r} r^{2} \right) +
  \frac{1}{r \sin\theta} \frac{\partial}{\partial r}
  \left( \rho v_{\phi} \right) \, = \, 0 \,\,\,\, ,
\end{equation}
and the conservation of the $r$ (radial) and $\phi$ (azimuthal)
components of momentum,
\begin{equation}
  \frac{\partial v_{r}}{\partial t} +
  v_{r} \frac{\partial v_{r}}{\partial r} +
  \frac{v_{\phi}}{r \sin\theta} \frac{\partial v_{r}}{\partial \phi}
  \, = \, \frac{v_{\phi}^{2}}{r} -
  \frac{1}{\rho} \frac{\partial P}{\partial r} +
  g_{r}^{\mbox{\footnotesize ext}} \,\,\,\, ,
\end{equation}
\begin{equation}
  \frac{\partial v_{\phi}}{\partial t} +
  v_{r} \frac{\partial v_{\phi}}{\partial r} +
  \frac{v_{\phi}}{r \sin\theta} \frac{\partial v_{\phi}}{\partial \phi}
  \, = \, - \frac{v_{r} v_{\phi}}{r} -
  \frac{1}{\rho r \sin\theta} \frac{\partial P}{\partial \phi} \,\,\,\, ,
\end{equation}
where $\rho$ is the mass density, $v_{r}$ and $v_{\phi}$ are the $r$
and $\phi$ components of the velocity, and $t$ is the time.
The $\theta$ (latitudinal) component of the momentum conservation
equation is assumed satisfied in the equatorial plane by the trivial
solution $v_{\theta} (\theta = \pi /2) \, = \, 0$, i.e.\  no latitudinal
flow into or out of the computational domain,
with all partial derivatives in the $\theta$ direction
considered negligible.
The code also includes an equation for the conservation of energy,
but in all models presented here this is dominated by rapid radiative
processes, which keep the gas very nearly isothermal
with a constant wind temperature $T$ equal to the stellar effective
temperature $T_{\mbox{\footnotesize eff}}$.  We use a perfect gas law
equation of state to evaluate the pressure $P$.

As in Owocki, Cranmer, \& Blondin \markcite{OCB}(1994), the external radial
acceleration here includes gravity and radiative driving by line
scattering, and
\begin{equation}
  g_{r}^{\mbox{\footnotesize ext}} \, = \,
 - \frac{GM_{\ast} (1 - \Gamma )}{r^2} +
  g_{r}^{\mbox{\footnotesize lines}} \,\,\,\, .
  \label{eq:geff}
\end{equation}
Here $G$ and $M_{\ast}$ are the gravitational constant and stellar
mass, and $\Gamma$ ($= \kappa_{e} L_{\ast} / 4 \pi GM_{\ast} c$)
is the Eddington factor that accounts for the reduction in effective
gravity by outward radiation pressure on electrons.
We evaluate the
line-scattering acceleration $g_{r}^{\mbox{\footnotesize lines}}$
in the local Sobolev \markcite{S60}(1960) approximation.
This suppresses the wind's strong line-driven instability
(Owocki \& Rybicki \markcite{OR84}1984), as it is not
currently feasible to incorporate this inherently nonlocal effect in
multidimensional hydrodynamic models.

For one-dimensional winds, the Sobolev line force per unit mass can
be parameterized as
\begin{equation}
  g_{r}^{\mbox{\footnotesize lines}} \, = \,  k f
  \left( \frac{1}{\kappa_{e} v_{\mbox{\footnotesize th}} \rho}
  \left| \frac{\partial v_{r}}{\partial r} \right| \right)^{\alpha}
  \frac{GM_{\ast}\Gamma}{r^2} \,\,\,\, ,
  \label{eq:glines}
\end{equation}
where the Castor, Abbott, \& Klein \markcite{CAK}(1975,
hereafter CAK) parameters
$\alpha$ and $k$ are related to the slope and normalization of the
assumed power-law ensemble of lines.
The constant $k$ is defined in terms of the electron scattering
coefficient $\kappa_e$ and a fiducial ion thermal speed
$v_{\mbox{\footnotesize th}}$ (Abbott \markcite{A82}1982).
The stellar radiation field is modeled here by a spherical star
with no limb darkening, and
$f$ is the finite-disk correction factor used
by Friend \& Abbott \markcite{FA86}(1986) and Pauldrach, Puls,
\& Kudritzki \markcite{PPK}(1986).
We ignore the effects of rotational oblateness and
gravity darkening on the full {\em vector} radiative force (Cranmer
\& Owocki \markcite{CO95}1995), which we take to be purely radial.
For simplicity, we also
do not use Abbott's \markcite{A82}(1982) added ionization-balance
parameterization of the line force,
and set his exponent $\delta =0$.

The Sobolev approximation assumes a monotonically accelerating
velocity field, but this is not guaranteed in the time-dependent
simulations presented below.
Winds with nonmonotonic velocities have {\em nonlocal} line forces,
since multiple resonance surfaces can create additional attenuation
of the stellar flux (Rybicki \& Hummer \markcite{RH78}1978;
POF\markcite{POF}).
We do not treat this nonlocal coupling directly, but we have compared
CIR models using an upper limit (unattenuated) and a lower limit
(strongly attenuated) for the line force in multiply-resonant regions
of the wind; we find the dynamics to be quite similar in both limits.
The upper limit, which we use in all models presented below, involves
taking the absolute value of the radial velocity gradient
$| \partial v_{r} / \partial r |$ in equation (\ref{eq:glines})
and in the finite disk factor $f$.
The lower limit assumes decelerating flows
($\partial v_{r} / \partial r < 0$) receive the same small force
contribution from nonradial rays as a flow that is not accelerating at all
($\partial v_{r} / \partial r =0$).

\subsection{Local Radiative Force Enhancement}

We induce azimuthal structure in our models by varying the Sobolev
line force over a localized ``star spot'' in the lower wind.
Since this force is directly proportional to the stellar flux
($L_{\ast} / 4 \pi r^{2}$), increasing or decreasing
$g_{r}^{\mbox{\footnotesize lines}}$ over a small area is
operationally equivalent to assuming
a bright or dark region of the photosphere.
Note, however, that by modulating the radiative force in this manner
we do {\em not} mean to literally propose the existence of
strong flux-varying spots on early-type stars.
We merely use this simple method to perturb the wind in lieu of more
definite knowledge about the physical cause(s) of surface
inhomogeneities.
Because the line driving grows weaker as one moves deeper into the
subsonic wind and photosphere, the force enhancement is essentially
confined to the transonic and supersonic wind, obviating the need
to model a perturbed stellar atmosphere.

The induced variation in the line force is assumed to have a specified
radial and azimuthal dependence which remains fixed to the stellar surface,
and thus rotates through the computational domain with the star.
The force enhancement is a function of radius $r$ and a
corotating azimuthal angle $\psi$,
\begin{equation}
  \psi \, = \, \phi - \Omega t \,\,\,\, ,
\end{equation}
with $\Omega \equiv V_{\mbox{\footnotesize eq}}/ R_{\ast}$
the star's constant rotational angular velocity.
The perturbed line force has the form
\begin{eqnarray}
  g_{r}^{\mbox{\footnotesize lines}} (r, \psi) &=&
   g_{0} (r) + \delta g (r, \psi) \nonumber \\
  &=& g_{0} (r) \left\{ 1 + A \,
   \gamma (r) \exp \left[ - \left( \psi - \psi_{0} \right)^{2}
   / \sigma^{2} \right] \right\} \,\,\,\, ,
   \label{eq:delg}
\end{eqnarray}
where $g_{0}$ is the unperturbed Sobolev
force (eq.~[\ref{eq:glines}]), $\psi_{0}$ is the azimuthal position
of the center of the spot, and $A$ is its dimensionless amplitude.
We will refer to ``dark spots'' as those with $-1 < A < 0$ and
``bright spots'' as those with $A>0$.
The azimuthal variation of the force is here assumed to be Gaussian
about $\psi_{0}$, with full width at half maximum (FWHM) given by
$\Phi \equiv 2 \sigma \sqrt{\ln 2}$.
The radial modulation $\gamma (r)$ is constrained by the geometrical
extent of the spot.
Close to the star, the spot is all that can be seen, and
so $\gamma (r \rightarrow R_{\ast}) \rightarrow 1$;
far from the star, the spot only represents a fraction
of the observed stellar disk, and
so  $\gamma (r)$ approaches
a small, but constant value as $r \rightarrow \infty$.

For a field point directly over ($\psi = \psi_{0}$) a {\em circular}
flux enhancement with angular diameter $\Phi$,
the radial function $\gamma (r)$ can be derived analytically from the
the normalized residual flux,
\begin{equation}
  \frac{{\cal F} - {\cal F}_{0}}{{\cal F}_{0}} \, = \,
  A \frac{r^2}{R_{\ast}^{2}} \int \! \int D (r, \mu' , \phi' )
  \, \mu' d \mu' d \phi'  \, = \, A \gamma (r) \,\,\,\, ,
\end{equation}
where ${\cal F}$ and ${\cal F}_{0}$ are the total and unperturbed
fluxes.
The amplitude $A$ takes into account the relative magnitude of
the spot's ``residual effective temperature,'' and is equivalent to
$(T_{\mbox{\footnotesize spot}}^{4} - T_{0}^{4})/T_{0}^{4}$.
The area integral is taken over a solid angle centered about the
$z$-axis, with angles $\theta' = \cos^{-1} \mu'$ and $\phi'$ measured
from the field point in the wind at radius $r$, and the residual
limb darkening function $D$
set to zero for rays not intercepting the spot.
Thus, for simple linear limb darkening,
\begin{equation}
  \gamma (r) \, = \, \frac{2 \pi r^{2}}{R_{\ast}^{2}}
  \int_{\mu_{0}(r)}^{1} \frac{1}{4 \pi} \left( 2 + 3
  \sqrt{ \frac{\mu'^{2} - \mu_{\ast}^{2}}{1 - \mu_{\ast}^{2}} }
  \right) \mu' d \mu' \,\,\,\, ,
  \label{eq:gamint}
\end{equation}
where $\mu_{\ast} \equiv [ 1 - (R_{\ast}^{2} / r^{2}) ]^{1/2}$
defines the stellar limb, and
\begin{equation}
  \mu_{0} (r) \, = \, \left\{
  \begin{array}{ll}
    \mu_{\ast},     & r \cos (\Phi /2) \le R_{\ast} \\
    \sqrt{1 - R_{\ast}^{2} \sin^{2} (\Phi /2) / S^{2}}, &
    r \cos (\Phi /2) > R_{\ast}
  \end{array} \right.
\end{equation}
defines the visible edge of the spot.
The distance $S$ from the field point to the edge of the spot is
$[ r^{2} + R_{\ast}^{2} - 2 r R_{\ast} \cos (\Phi /2) ]^{1/2}$.
We thus adopt the radial modulation function given by the analytic
integral of (\ref{eq:gamint}),
\begin{equation}
  \gamma (r) \, = \, \frac{1}{2} \left[ 1 + \left(
  \frac{1 - \mu_{0}^{2}}{1 - \mu_{\ast}^{2}} \right) -
  \left( \frac{\mu_{0}^{2} - \mu_{\ast}^{2}}{1 - \mu_{\ast}^{2}}
  \right)^{3/2} \right] \,\,\,\, ,
  \label{eq:gamm}
\end{equation}
which approaches unity as $r \rightarrow R_{\ast}$ and approaches
a constant value of
$[1 + \sin^{2} (\Phi /2) - \cos^{3} (\Phi /2) ] /2$ as
$r \rightarrow \infty$.
Although this residual flux integral is also able to provide oblique
($\psi \neq \psi_{0}$) and nonradial components of the flux
enhancement of a star spot, we restrict our present models to the
explicit {\em radial} perturbation given by equations
(\ref{eq:delg}) and (\ref{eq:gamm}).
Again we emphasize that we do not mean to model in detail an
actual star spot, but are only using the spot-like force
enhancement as a convenient way to perturb the wind base.

Figure~1 shows contours of the force enhancement in the equatorial
plane for a spot with $\Phi = 20\arcdeg$,
as well as wind ``streaklines'' for various stellar rotation
speeds, as discussed further below (\S~3.3).
Note that the spot significantly affects only a relatively small
area of the wind:  an azimuthal extent of $\sim 2 \Phi$ and a
radial extent of about a stellar radius.
This allows several spots to be superposed on the stellar surface
without any appreciable overlap in their force enhancements.
Following the empirical arguments of Kaper \& Henrichs \markcite{KH94}(1994),
who suggest a variable dipole magnetic field as the seed of large-scale
wind structure, we place {\em two} spots separated by $180\arcdeg$
on our model stars.

Before examining how a rotating wind responds to the localized force
enhancement, it is instructive to see how a non-rotating wind is
affected.
Figure~2 shows the radial velocity and density (directly over the spot)
at a reference radius of $10 R_{\ast}$, well beyond the region of
significant direct force enhancement, as a function of amplitude $A$.
The general trend is for a ``bright'' spot to increase the local
mass loss and thus increase the density of the wind near the star.
Further out in the wind, where the effect of
the spot drops off, the radiative force cannot accelerate the
higher density material as strongly,
so it approaches a {\em lower} terminal speed
than in the unperturbed wind.
Conversely, ``dark'' spots decrease the mass loss in the
surrounding wind and thus allow the less dense material to be
accelerated more strongly, leading to a much {\em higher} terminal speed.
The three sets of data in Figure~2 correspond to:
(1) one-dimensional, finite-disk, ``modified CAK'' (mCAK) solutions
with a realistic critical point analysis and numerical integration,
(2) two-dimensional hydrodynamical models using VH-1, and
(3) a simple analytic fit to the data.
The one-dimensional mCAK models contain the radial spot modulation
$\gamma (r)$, but no information about neighboring streamline divergence
or convergence.
For spots with $A \gtrsim 0.6$, the one-dimensional models cease to
have steady-state solutions that reach to infinity because too much mass
is driven off the star to be accelerated beyond its gravitational escape
velocity.
The two-dimensional models can drive more mass to infinity because
the density is reduced by a slight azimuthal expansion, which leads to
a faster-than-radial divergence of flow tubes (see, e.g., MacGregor
\markcite{M88}1988).

The simple fit to the velocity and density variation in Figure~2
depends on only one free parameter, and makes use of the approximate
dependence of the CAK mass loss rate on an arbitrary force multiplier
(see Cranmer \& Owocki \markcite{CO95}1995, eq.\  24).
Directly over the spot,
\begin{equation}
  \dot{M} \, \approx \, \dot{M}^{(0)} (1 + A)^{1/ \alpha} \,\,\, ,
\end{equation}
where $\dot{M}^{(0)}$ is the unperturbed mass loss rate, and we assume
that the CAK\markcite{CAK} critical point $r_{c}$, where the mass flux
is determined,
is close enough to the star that $\gamma (r_{c}) \approx 1$.
In a one-dimensional steady state, the mass conservation equation is
integrated in the usual way to obtain
$\dot{M} = 4\pi \rho v_{r} r^{2}$,
and this provides a relation between the velocity and density at a
given radius.
We thus define fitting functions which obey this multiplicative
constraint:
\begin{equation}
  \rho \, \approx \, \rho^{(0)} (1+A)^{s/ \alpha}  \,\,\, ,
  \label{eq:nonrot}
\end{equation}
\begin{equation}
  v_{r} \, \approx \, v_{r}^{(0)} (1+A)^{(1-s)/ \alpha}  \,\,\, ,
  \label{eq:nonrotv}
\end{equation}
where we have found the best fit value of $s=1.77$ for our dashed-line
fits in Figure~2.

\subsection{Numerical Specifications}

Let us next describe some of the details of our numerical discretization,
boundary conditions, and initial conditions.
We specify flow variables on a fixed two-dimensional spatial mesh
in radius $r$ and azimuthal angle $\phi$.
In our standard models we use 200 radial zones, from $R_{\ast}$
to $30 R_{\ast}$, with the zone spacing concentrated near the stellar
base where the flow gradients and spot enhancements are strongest.
The radial spacing starts at the lower boundary with
$\Delta r = 0.002 R_{\ast}$, then increases by 3\%
per zone out to a maximum of $\Delta r = 0.82 R_{\ast}$ at the
outer boundary.
The azimuthal mesh contains 160 constantly-spaced zones, ranging
from $0\arcdeg$ to $180\arcdeg$ with a spacing of 1$\fdg$125.
Limited test runs with double the resolution in radius and azimuth
showed some correspondingly greater detail in wind fine structure
(e.g., shocks and radiative-acoustic waves), but overall the results
were qualitatively similar to those for the standard resolution.

We specify the boundary conditions in our numerical method in two
phantom zones beyond each edge of the grid.
The azimuthal boundaries at $\phi = 0\arcdeg, 180\arcdeg$ are periodic,
ensuring symmetry in the full equatorial plane.
At the outer radial boundary, the wind is invariably supersonic
outward, and so we set the flow variables in the outer phantom zones
by a simple constant-gradient extrapolation.
The lower radial boundary of the wind is somewhat more problematic,
and we use the boundary conditions described by Owocki et
al.\  \markcite{OCB}(1994):
constant-slope extrapolation for $v_{r}$, rigid rotation for
$v_{\phi}$, and a fixed base density $\rho_{B}$.
Because the mass loss rates of line-driven winds are determined from
the equations of motion alone, we are able to specify an appropriate
``photospheric'' density
that yields a stable, subsonic boundary outflow
(see also Owocki, Castor, \& Rybicki \markcite{OCR}1988).

The time-dependent hydrodynamical
method requires a reasonable initial condition
to be specified over the entire grid at time $t=0$.
For this we relax an analytically derived mCAK model to a steady state
on a one-dimensional numerical grid, and then copy this onto the full
two-dimensional mesh.
This ensures that any time dependence results only from the
induced force perturbations.
The models are stepped forward in time at a fixed fraction (0.25)
of the standard Courant-Friedrichs-Lewy time step.
Because the radiative force enhancement, switched
on at $t=0$, only varies
in time by corotating with the stellar surface, the wind responds
by forming an outwardly moving ``front,'' behind
which the wind has settled to a rotating steady state.
The dynamical flow time for gas to radially cross the computational grid
is approximately $2 \times 10^{5} \, \mbox{s}$
for our unperturbed initial state wind.
Typically we find that models perturbed by star spots settle to a
corotating steady state within two
dynamical flow times, and so
we plot all models at
$t= 4 \times 10^{5} \, \mbox{s}$.

\section{Numerical Results}

Because the DAC phenomenon is primarily observed in O-star winds, we choose
to center our study on a standard model of
the O4f supergiant $\zeta$~Puppis.
Specifically, we take
$M_{\ast}=60 M_{\odot}$,
$R_{\ast}=19 R_{\odot}$,
$L_{\ast}=8 \times 10^{5} L_{\odot}$, and
$T_{\mbox{\footnotesize eff}} = 42,000 \, \mbox{K}$
(see, e.g., Howarth \& Prinja \markcite{HP89}1989;
Kudritzki et al.\  \markcite{Ke92}1992).
The measured rotational $V_{\mbox{\footnotesize eq}} \sin i$ for
$\zeta$~Puppis is $230 \, \mbox{km s}^{-1}$,  which we take
for the equatorial rotation velocity of our standard model.
We neglect the small ($\sim$7\%) oblateness induced by this
degree of rotation, which corresponds to a Roche equipotential
surface rotating at 63\% of its critical angular velocity.
We assume an isothermal wind of temperature
$T_{\mbox{\footnotesize eff}}$, corresponding to a sound speed
$a= 24 \, \mbox{km s}^{-1}$, and use the line-driving constants
$\alpha = 0.60$ and $k=0.15$ (see eq.\  [\ref{eq:glines}]).
In one-dimensional mCAK models, these result in a terminal velocity
$v_{\infty} = 2580 \, \mbox{km s}^{-1}$ and a mass flux
$\dot{M} = 3.28 \times 10^{-6} \, M_{\odot} / \mbox{yr}$.
The base density for our subsonic lower boundary condition is
$\rho_{B} = 6 \times 10^{-11} \, \mbox{g cm}^{-3}$.

The localized ``star spot'' radiative force enhancement described above
depends primarily on two quantities:  the amplitude $A$ and the
azimuthal full width $\Phi$.
These, together with the equatorial rotation velocity
$V_{\mbox{\footnotesize eq}}$, are the three free parameters we vary
in our study of non-axisymmetric structure formation.
Table~1 outlines the input parameters and several output quantities
(to be discussed below) for the models we computed.
Models~1 and 2 are standard ``bright'' and ``dark'' spot models,
with $A= +0.5$ and $-0.5$, and they represent a basis to explain the
general hydrodynamical phenomenon of stream interaction.
The subsequent models in Table~1 are intended to confirm our
understanding of the physics of CIR formation, and are
discussed below in a more limited fashion.

\subsection{Standard Bright Spot:  Model 1}

Figure~3 shows gray-scale plots for the density, radial velocity,
azimuthal velocity, and radial Sobolev optical depth in Model~1,
{\em normalized to the unperturbed wind.}
To ease comparison with other models,
the azimuthal coordinates here have been incremented by a
constant factor to align the peak of the spot ($\psi_{0}$)
with the center-line or $x$-axis of the diagram.
Note the expected tendency for a bright spot to create higher
density and lower radial velocity.
However, the {\em azimuthal} velocity only differs by a small subsonic
amount from the unperturbed
angular-momentum-conserving form
$v_{\phi}^{(0)} (r) =
V_{\mbox{\footnotesize eq}} R_{\ast} /r$.
This demonstrates the almost purely {\em radial}
effect of the spot enhancement.

Most of the corotating structure from the spot settles onto nearly
constant spiral ``streaklines'' in the wind.
In the present models, streaklines are equivalent to flow streamlines in
the star's rotating frame of reference.
Figure~4 compares streamlines and streaklines computed for Model~1.
By numerically integrating the kinematic relation
\begin{equation}
  \frac{r \, d\phi}{dr} \, = \, \frac{v_{\phi} (r,\phi) - r \Omega_{F}}
  {v_{r}(r,\phi)} \,\,\, ,
  \label{eq:streak}
\end{equation}
from a locus of points on the stellar surface
spaced evenly in $\phi$, one can alternatively compute either
streamlines in the inertial reference frame,
with $\Omega_{F} =0$, or
streaklines in the rotating reference frame,
with $\Omega_{F} =\Omega$.
Areas with a higher (lower) concentration of streaklines correspond
to regions of relative compression (rarefaction), though
not all density variations are reflected in the streaklines.
The dashed lines in Figure~4 show the streamline and streakline
originating directly over the spot.
The streamline appears nearly radial, but careful inspection
shows that it has a modest ($18\arcdeg$) prograde deflection,
resulting from corotation of the relatively slow wind outflow near
the surface.
Through most of the wind, however, the streamlines are
close to radial, and this allows one to qualitatively
interpret the azimuthal coordinate as a {\em time} dimension.
For any streamline at a fixed value of $\phi$, which is
intercepted by different streaklines as the (corotating steady
state) system sweeps by, the two-dimensional hydrodynamics becomes
effectively one-dimensional, but now truly time dependent.
This concept allows us to understand the ``spread out'' CIR structure
in terms of a simpler model of radial wave or shock propagation.
Hundhausen \markcite{H73}(1973) modeled solar-wind CIR formation and evolution
in one dimension using this approximation.

We can disentangle the actual patterns of high density, low velocity, and
more optically thick CIR structure by examining several
causally-connected regions of this model:
\begin{description}
\item[I.] {\bf Direct Enhancement:}
Close to the stellar surface, the Gaussian-shaped spot increases the
mass flux and wind density over a limited ($r \lesssim 2 R_{\ast}$)
region near the star.
This enhanced-density patch is slightly deformed by rotation from the
contours shown in Figure~1, but is essentially equivalent.
The density increases over the spot by a maximum factor of $\sim$2.6,
only slightly smaller than that predicted by the non-rotating analysis
(see eq.~[\ref{eq:nonrot}]).
This region also shows considerable azimuthal spreading in $v_{\phi}$
as the wind begins to adjust to the presence of the spot.
\item[II.] {\bf Prograde Precursor:}
Just ahead of the spot ($\psi > \psi_{0}$), a small fraction of the
enhanced higher-density wind is able to ``leak out'' azimuthally and
settle onto a set of relatively unperturbed streaklines.
The density in this feature is only enhanced by a factor of $\sim$1.2,
indicating that it comes from material in the prograde tail of the
Gaussian distribution.
It becomes isolated from the direct spot enhancement at a relatively
large distance from the star ($r \approx 6.4 R_{\ast}$)
after the CIR rarefaction (IV) has appeared
between it and the CIR shock (III).
\item[III.] {\bf CIR Compression:}
The low radial velocity wind from the center of the spot curls around on
more tightly-wound streaklines than the surrounding unperturbed wind, and
these streams begin to interact at a finite radius from the star
($r \approx 1.6 R_{\ast}$).
Alternately, in the above one-dimensional interpretation, the slow
stream can be considered equivalent to a radially-extended Gaussian
``wave packet'' which nonlinearly steepens as the fast mean wind
begins to overtake it.
The result of this collision of fast and slow streams is a corotating
weak shock compression (the CIR) which, because it is driven by ram
pressure from the mainly unperturbed
wind, propagates out at very near
the unperturbed wind velocity.
Because the flow is isothermal, we do not see a separation into a distinct
forward and reverse shock pair, as is observed in the more nearly adiabatic
solar wind.
\item[IV.] {\bf CIR Rarefaction:}
Ahead of the nonlinear shock the streaklines fan out and form a
lower-density rarefied region.
The formation of this rarefaction is mandated by mass flux
conservation, and the radial velocity correspondingly peaks slightly
above its unperturbed value here.
Because the density in this feature never dips too far below the
unperturbed density ($\min [\rho / \rho^{(0)}] \approx 0.82$),
the rarefaction propagates out at nearly the same velocity as the
CIR compression.
\item[V.] {\bf Radiative-Acoustic ``Kink:''}
In a purely hydrodynamical wind, the radial CIR shock structure is the
sole result of the nonlinear steepening of the initial enhancement.
Any nondissipative signals propagating in the rest frame
of the wind (at characteristic speeds $\pm a$) are limited to the
relatively undisturbed lateral (nonradial) direction.
In a line-driven wind, however,
Abbott \markcite{A80}(1980) and Rybicki et al.\  \markcite{Re90}(1990)
found that
large spatial-scale linear perturbations propagate in the radial
direction at modified ``radiative-acoustic'' characteristic speeds.
Abbott \markcite{A80}(1980) derived
\begin{equation}
  C_{\pm} \, = \, -\case{1}{2} U \pm \sqrt{ \left( \case{1}{2} U
  \right)^{2} + a^{2} }
\end{equation}
for radial modes, where $U \equiv \partial (
g_{r}^{\mbox{\footnotesize lines}} ) / \partial ( \partial v_{r} /
\partial r )$ and $C_{\pm}$ reduces to the purely acoustic case if
$U=0$.
In most of the wind, though, $U \gg a$, and the outward (positive root)
solutions are subsonic, and the inward (negative root) solutions are
supersonic.
In Model~1 we see both an acoustic lateral mode (spreading in
$v_{\phi}$ at large radii; Figure~3c) and a nonlinear analog of the
inward radiative-acoustic mode, which propagates
slowly outward in the star's frame ($0 < v_{r} + C_{-} < v_{r}$)
as a weak discontinuity, or ``kink,'' in the
radial velocity gradient.
Because of its slow propagation (more tightly-wound streaklines)
this feature eventually collides with the CIR rarefaction from the
other spot at a radius of $\sim$13.5~$R_{\ast}$ and ceases to exist.
\end{description}

We trace these five features
in Figure~5, which is a
close-up of the density gray-scale shown in Figure~3a.
Unique tracks were found by searching for local extrema (in radius)
of various quantities, and following contiguous patterns around in
azimuth.
The direct spot enhancement (I) appears at the stellar surface
as a local maximum in the
normalized density $\rho / \rho^{(0)}$, and collides with the
CIR/kink pair of features (III and V) at a radius of
$\sim$3.9$R_{\ast}$.
These ``bifurcated'' extrema are found by tracking local minima and
maxima in the radial velocity, as shown in Figure~6 below.
They appear at a relatively small radius $r_{L}$ (see Table~1),
where the spot perturbation is still linear.
The remaining precursor/rarefaction pair of features (II and IV)
correspond to other local minima and maxima in the normalized density,
and they appear further out (at a larger radius $r_{NL}$) where
the disturbance has definitely steepened into a nonlinear shock.
It is interesting that the CIR compression and rarefaction do not
form together at the same point, but this is understandable, since the
latter can be considered an effect or response of the former.

To get an indication of which wind structures should yield the most
prominent signatures in observed line profile variations, let us examine
the radial Sobolev optical depth,
\begin{equation}
  \tau_{r} (r,\phi) \, \equiv \, \frac{\kappa_{L} \, \rho (r,\phi)
  \, v_{\mbox{\footnotesize th}}}{\left| \partial v_{r} (r,\phi) /
  \partial r \right|} \,\,\, .
  \label{eq:tausob}
\end{equation}
The gray-scale plot in Figure~3d shows the changes in the optical depth
relative to the mean, unperturbed wind.
Since both the line absorption coefficient $\kappa_{L}$ and
the ion thermal speed $v_{\mbox{\footnotesize th}}$ are assumed to have
the same constant values in both the mean and perturbed flow, all variations
here stem from changes in the ratio of density to velocity gradient.
Somewhat surprisingly, the regions of strongest optical depth enhancement
occur not within the dense CIR compression (feature III), but rather within
the relatively shallow velocity gradient region after the Abbott kink
(feature V).

Figure~6 plots the radial variation of velocity and density from selected
slices of constant azimuthal angle $\phi$.
For this corotating steady state, the changing features in these line plots
also indicate the time evolution of structure at fixed azimuths.
This allows us to follow the outward propagation of both
the CIR density enhancement (local minima in velocity) and the
trailing radiative-acoustic kink (local maxima in velocity).
In the high-density CIR, the wind is either strongly accelerating or
decelerating, so both the numerator and the denominator in the Sobolev optical
depth (eq.~[\ref{eq:tausob}]) are enhanced, resulting in very little
net increase.
Just outward from the kink, however, the density is nearly unperturbed,
while the velocity gradient is much shallower, implying a large increase
in the Sobolev optical depth.
In synthetic line profiles (see \S~4.2 and Figure~11a), this near
plateau produces a distinct absorption feature quite similar to
slowly evolving DACs.
As the (corotating) steady-state structure rotates in front of the
observer's line of sight, material flowing {\em through} the kink appears
at the velocities of the local maxima in Figure~6,
but the evolution of the feature is governed by the
radiative-acoustic mode propagation, which leads to an
apparent acceleration that is much slower than the
actual acceleration of the wind material.

Ahead of the CIR, between the compressive
density maximum and the rarefied minimum, there is a region
of high acceleration that arises from the
prograde edge ($\psi > \psi_{0}$) of the Gaussian spot, which
steepens into a monotonic sawtooth structure connecting
the shock with the lower-density unperturbed wind.
This region contains a lower net optical depth than the unperturbed
wind, implying a relative {\em lack} of absorption in synthetic line
profiles.
This effect is only slightly weaker than the enhanced absorption due
to the plateau, suggesting that isolated patches of extra
absorption (DACs) may be difficult to model theoretically without a
corresponding lower optical depth feature (with apparent relative
``emission'').

\subsection{Standard Dark Spot:  Model 2}

A wind perturbed by a locally decreased radiative force
($A = -0.5$) produces a lower-density, high-speed stream, and thus
settles to a steady state on the computational grid faster than
a model with slow streams.
Figure~7 shows gray-scale plots for the density, radial velocity,
azimuthal velocity, and radial Sobolev optical depth for Model~2,
normalized in the same way as in Figure~3.
Although the dark spot produces an extremely rarefied wind, a
high-density CIR forms (on the leading edge of the perturbation) where
the high velocity stream collides with the slower unperturbed wind.
The corotating structure present in Model~2 is qualitatively simpler
than that in Model~1.
The CIR/kink pair of features initially appears at
$r_{L} \approx 2.2 R_{\ast}$, and advects smoothly throughout the wind.
There is no analog to the second pair of features (starting further
out at $r_{NL}$) in this model.

For slices of constant $\phi$, Figure~8 plots the radial dependence
of the velocity and density.
The contrast with the slow structure in
Figure~6 is apparent.
Note that the back-propagating radiative-acoustic kink is also
present in this model, comprising the left edge of the flat-topped
velocity peaks.
However, since the kink here is a reaction to the forward-steepened
shock, it forms within the fast and rarefied upstream region, and thus
does not contribute strongly to the Sobolev optical depth of the
corotating feature.
The high density CIR also does not have an enhanced opacity
because of the steep velocity gradients near the shock, as in Model~1.
Indeed, Figure~7d shows that most of the highest optical depth
material comes from the unperturbed wind, and that the high-speed
CIR should be mainly a source of {\em decreased} absorption.
The isolated ``clumps'' of highest optical depth in Figure~7d
(and in Figure~3d) are artifacts of the finite differencing used
to compute the radial velocity gradient, and do not significantly
affect the volume-integrated quantities used in constructing
line profile diagnostics.

The CIR structure in Model~2 appears very similar to that seen in
hydrodynamic models of the solar wind, e.g., note the resemblance
between Figure~8 and Figures~2 and 3 of Hundhausen \markcite{H73}(1973).
The two major differences between solar wind high-speed streams and
those in our model are:
(1) the distinct forward and reverse adiabatic shocks in the former and
(2) the back-propagating radiative-acoustic kink in the latter.
Despite this similarity, all subsequent models in our parameter study
use the bright spot of Model~1, which promises to simulate better the
slow DACs in early-type stellar winds.

\subsection{Variation of Spot Amplitude, Width, and Stellar
Rotation Velocity}

Let us now examine the effect of changing various spot and wind parameters.
To provide a basis for understanding the full hydrodynamical calculations,
we can estimate the expected effects in terms of a simple wind
streakline picture (see Figure~4).
Note that the shape of these streaklines can be
well approximated by neglecting both the wind's acceleration and
the angular-momentum-conserving azimuthal velocity, which nearly ``cancel
each other out'' when computing streakline deflection.  Thus, the
angular deflection is given in this approximation by the Archimedean
spiral relation,
\begin{equation}
  \phi - \phi_{0} \, \approx \, - \frac{\Omega}{v_{\infty}} \,
  (r - r_{0}) \,\,\,\, .
  \label{eq:archimedes}
\end{equation}
Mullan \markcite{M84a}(1984a)
used this to estimate the interaction radius $r_{i}$ between
a fast and slow stream initially set apart on the stellar surface by
a given azimuthal
separation $\Delta \phi$,
\begin{equation}
  \frac{r_{i}}{R_{\ast}} \, = \, 1 +
  \frac{\Delta \phi}{V_{\mbox{\footnotesize eq}}} \left(
  \frac{v_{\infty}^{(f)}   v_{\infty}^{(s)}}
       {v_{\infty}^{(f)} - v_{\infty}^{(s)}} \right) \,\,\, .
  \label{eq:ri}
\end{equation}
Here
$v_{\infty}^{(f)}$ and $v_{\infty}^{(s)}$ are the terminal velocities
of the fast and slow streams.
Table~1 contains this simple prediction for $r_{i}$ for all the
hydrodynamical models in our parameter study.
We assume $\Delta \phi = \Phi$, and for bright-spot models, we
take the fast stream to be the unperturbed wind, and the slow stream
to have $v_{\infty}^{(s)}$ given by equation~(\ref{eq:nonrotv}).
Conversely, for the dark-spot model the fast stream is given by
equation~(\ref{eq:nonrotv}) and the slow stream is unperturbed wind.

Models~3A, 3B, and 3C vary the spot amplitude $A$, and thus the
direct enhancement in density and velocity over the spot (see Table~1).
The resulting structure looks qualitatively similar to that of Model~1,
but the fast/slow stream interaction takes place at different radii.
Figure~9a traces the corresponding CIR compression and
radiative-acoustic kink features for these models.
As expected, CIRs form further out when there is a
smaller discrepancy between the fast and slow stream speeds,
but the linear minima and maxima in velocity first appear
(at $r_{L}$) much closer to the star than the simple interaction
analysis above predicts.
The ``nonlinear'' radius $r_{NL}$, where the CIR rarefaction
branches off from the prograde dense precursor, is relatively
{\em constant} with $A$, indicating a qualitatively different
formation mechanism from the compression/kink features.

Figure~10 shows how the velocity law of the kink and CIR shock varies
with spot amplitude, with each model rotated in azimuth to line up
similar features.
For the smallest values of $A$ (0.01, 0.1), no shock has yet formed,
so both features appear ``kink-like,'' propagating nearer
to the slow radiative-acoustic mode speed than to the mean wind speed
(see Figure~9a).
The shock steepening is evident for larger values of $A$, and the most
extreme model ($A=2.50$) shows considerable slowing of the dense
CIR resulting from the inverse dependence of the radiative force with
density (eq.~[\ref{eq:glines}]).
Our canonical Model~1 CIR, then, seems just on the verge between both
``slowing'' mechanisms, as well has having a shape between the
low-$A$ weak discontinuities and the high-$A$ shocks.
In all cases, however, the decelerating plateau-region ahead of the
kink has the same characteristic (negative) acceleration, indicating
that the enhanced optical depth at this radius may not vary strongly
with spot amplitude.

Models~4A, 4B, and 4C vary the azimuthal spot width $\Phi$, and Figure~9b
shows their CIR compression and kink features.
Increasing or decreasing the full-width $\Phi$ simply alters the
spatial scale of the interactions, and, in the limit of very small
spots (where the star's sphericity can be neglected) the CIR structures
seem self-similar with respect to an overall expansion factor.
Table~1 indicates an inverse relationship between $\Phi$ and the
maximum CIR density, and this can be understood heuristically by the
fact that when a given spot amplitude is spread over a larger area,
the collision of fast and slow streams is more diluted, and the
resulting shocks are not as strong.
Model~4A, with the smallest width $\Phi=10\arcdeg$, shows an apparent
reversal in this trend, and we suspect that some of the detailed
shock structure is under-sampled in our relatively coarse (in this
case) azimuthal grid.
Note that Mullan's \markcite{M84a}(1984a)
interaction radius $r_{i}$ agrees well with
the location of the nonlinear rarefaction/precursor feature $r_{NL}$
for these models.

Observations (Prinja \markcite{P88}1988; Henrichs et
al.\  \markcite{HKZ}1988)
suggest that the recurrence and acceleration times of DACs tend to
vary inversely with the projected equatorial rotation speed,
$V_{\mbox{\footnotesize eq}} \sin i$.
To examine how well our dynamical models might reproduce these observed
trends, Models~5A and 5B vary the equatorial stellar rotation velocity
$V_{\mbox{\footnotesize eq}}$.
Figure~9c shows their CIR compression and kink features.
The dominant effects are the overall variation of the spiral streaklines
(see, e.g., eq.~[\ref{eq:archimedes}]), and the
inverse centrifugal dependence of
$v_{\infty}$ on the rotation rate (Friend \& Abbott \markcite{FA86}1986).
The strength of the CIR shock decreases for more slowly rotating stars,
and we believe that this results from the variation of the streaklines
with respect to the star-spot enhancement (Figure~1).
Perturbed gas which rapidly advects out of the region of direct enhancement
receives less of a ``boost'' of extra density, and thus does not form
as strong a compression when interacting with the ambient wind.

\section{Synthetic Observational Diagnostics}

\subsection{SEI Line Profile Construction}

Ultraviolet P~Cygni lines are sensitive probes of the wind structure
of hot luminous stars, and it is important to model accurately
observed variations in their profile shape.
Here we use a multidimensional extension of the SEI (Sobolev with Exact
Integration) method of Lamers, Cerruti-Sola, \& Perinotto
\markcite{Le87}(1987) to
compute synthetic line profiles.
In this method, the source function is calculated locally in the wind
using the Sobolev escape probability approximation, and the emergent
flux profile is computed by numerically integrating the formal solution
to the transfer equation.
Bjorkman et al.\  \markcite{Be94}(1994)
discuss a two-dimensional extension of the
basic SEI algorithm, and our computational approach is similar in that
we do not yet treat the general case of nonlocal (or line doublet)
resonance coupling.
Our method, however, efficiently computes line profiles from an
arbitrary three-dimensional distribution of density and velocity,
for observers at arbitrary vantage points.

Following the notation of POF\markcite{POF}, we parameterize the
opacity of a model pure-scattering resonance line by defining
a dimensionless line-strength
\begin{equation}
  k_{L} \, \equiv \, \left(
  \frac{\dot{M} v_{\mbox{\footnotesize th}}}
  {4 \pi R_{\ast} v_{\infty}^{2}}
  \right) \kappa_{L} \,\,\, ,
\end{equation}
where $v_{\infty}$ and $\dot{M}$ are taken from the unperturbed model wind.
We assume the mass absorption coefficient
$\kappa_{L}$ is constant in radius, which is valid for
lines of interest in the dominant ionization stage of the wind, and at
least allows qualitative comparison for other lines.
We consider two representative cases:  a moderate unsaturated line
($k_{L}=1$) and a strong saturated line ($k_{L}=100$).

The local two-dimensional (solid angle) integrals required to obtain the
escape probability and core-penetration probability for the Sobolev
source function are computed using Romberg's successive refinement
algorithm.
The explicit form of these integrals is found in Lamers et
al.\  \markcite{Le87}(1987).
POF\markcite{POF} compare the use of the local Sobolev formalism
with a self-consistent multiple-resonance technique in structured wind
models, and find significant disagreement in the resulting line
profiles.
However, our CIR model winds are much less structured than the
one-dimensional instability models used by POF\markcite{POF},
with at most only two zones of nonmonotonic velocity
variation in the entire wind.
Further, here we concentrate on {\em residual} line profile
variability, which should be
less susceptible to consistency errors in the source function than
the actual line profile shape.

We perform the ``exact integration'' for the line flux using a
cylindrical $(p', \phi', z')$ coordinate system with the observer
oriented along the positive $z'$-axis at an infinite distance from
the origin.
The equation of radiative transfer is evaluated in differential form
along rays parallel to this axis, and along each ray the specific
intensity is integrated using second order implicit Euler differencing.
The resulting emergent intensities at the outer boundary of the
computational grid are then integrated by nested Romberg quadrature
in $p'$ and $\phi'$ to form the flux, and this process is repeated
for each frequency point in the total line profile.
We refer the reader to Lamers et al.\  \markcite{Le87}(1987) and
POF\markcite{POF}, who summarize
these integrals, but only apply them in the spherically symmetric
($\phi'$ independent) case.

We make two major approximations in our line profile construction
technique:
\begin{enumerate}
\item The three-dimensional wind structure is formed by interpolating
in latitude between the two-dimensional equatorial plane models and
the one-dimensional unperturbed polar wind.
For simplicity we assume the same Gaussian structure of the ``star
spots'' in latitude as in longitude, and apply it to the entire wind:
\begin{eqnarray}
  \rho (r,\theta,\phi) &=& \rho^{(0)} (r) \, [1 - E(\theta)] \, + \,
     \rho^{\mbox{\footnotesize (2D)}}(r,\phi) \, E(\theta)  \nonumber \\
  v_{r} (r,\theta,\phi) &=& v_{r}^{(0)} (r) \, [1 - E(\theta)] \, + \,
     v_{r}^{\mbox{\footnotesize (2D)}}(r,\phi) \, E(\theta) \\
  v_{\phi} (r,\theta,\phi) &=& v_{\phi}^{(0)} (r) \,
    [1 - E(\theta)] \, \sin\theta \, + \,
     v_{r}^{\mbox{\footnotesize (2D)}}(r,\phi) \, E(\theta)
     \, \sin\theta  \,\,\,\, , \nonumber
\end{eqnarray}
where $E(\theta) \equiv \exp [ -( \pi/2 - \theta)^{2} / \sigma^{2} ]$
(see eq.~[\ref{eq:delg}]).
As in the two-dimensional models, we retain our assumption that
$v_{\theta}=0$.
Note that the azimuthal velocity has an extra factor of $\sin\theta$
included to preserve angular momentum conservation out of the equatorial
plane, and this provides a latitudinal wind variation even for
{\em unperturbed} models.
\item
The total Doppler width of the Gaussian line profile contains both
thermal and microturbulent contributions,
\begin{equation}
  v_{D} \, \equiv \, \sqrt{
  v_{\mbox{\footnotesize th}}^{2} +
  v_{\mbox{\footnotesize turb}}^{2} } \,\,\,\, ,
\end{equation}
with the thermal speed set by $v_{\mbox{\footnotesize th}} = 0.28 a$,
as appropriate for CNO driving ions.
For simplicity, we assume a constant microturbulent velocity
$v_{\mbox{\footnotesize turb}}= 100  \, \mbox{km s}^{-1}$, though
better line-profile fits have been obtained by assuming this
varies in proportion to the mean wind velocity
(Haser et al.\  \markcite{He95}1995).
Though phenomenological, this use of a microturbulent velocity allows
realistic line-profile synthesis with a minimum of free parameters.
It also compensates for the suppression here of the small-scale instability,
which one-dimensional, nonlocal simulations have shown to result in
many of the same observational signatures as microturbulence
(POF\markcite{POF}; Owocki \markcite{O94}1994).
\end{enumerate}

\subsection{Time Variability in Dynamical Models}

We produce time-variable P~Cygni line spectra by positioning an
``observer'' in the equatorial plane
at successive azimuthal angles $\phi$ with respect to the
two-dimensional models.
For the standard rotation velocity used in Models 1, 2, 3, and 4,
the line profile variability repeats with a period
\begin{equation}
  \Pi \, = \, \frac{\pi R_{\ast}}{V_{\mbox{\footnotesize eq}}}
  \, = \, 2.091 \, \mbox{days} \,\,\, .
\end{equation}
Of course, since spiral streakline structures in the models often
subtend more then 180$\arcdeg \,$ of azimuth, continuous DAC-like signatures
can exist for times longer than this period.
We arbitrarily define $t=0$ when the observer is positioned directly over
the center of one of the photospheric spot perturbations.

As emphasized by Lamers \markcite{L94}(1994),
we find that most of the line profile
variability occurs in the absorption column of the wind
($0 \leq p' \leq R_{\ast}$),
with comparatively little variation in the
larger emission volume ($p' > R_{\ast}$).
Because the flux integration over this emission
volume dominates the CPU time in our SEI line-profile synthesis,
we only computed full, variable emission-volume profiles for
a few selected test cases, namely for
observers at four equally-spaced azimuthal
angles ($\phi = 0\arcdeg, 45\arcdeg, 90\arcdeg, 135\arcdeg$) for Model~1,
as well as at one arbitrary azimuthal angle for an unperturbed
$\phi$-independent wind model.
For an unsaturated ($k_{L} = 1$) line,
we found that the perturbations in the absorption-column flux
reach as high as 47\% of the continuum level, whereas those in
emission-volume flux never exceed 1.9\%.
Since these latter variations would only be
marginally observable with IUE signal-to-noise
ratios of 20 to 40, we neglect the perturbed emission
volume when computing subsequent line profiles on a finer time-
and velocity-sampled grid.

Figures~11a and 11b show the absorption-column line profile variability
for Model~1, repeated over three data periods
(1.5 rotation periods) to emphasize the
rotationally-modulated structure.
Following the standard observational convention, the gray-scale is
normalized by a ``minimum absorption'' (maximum flux)
template, constructed independently at each line velocity.
This choice contains the implicit bias that the variability
takes the form of extra {\em absorption,} which is
only partially appropriate for Model~1.
The unsaturated ($k_{L}=1$) line exhibits definite DACs that apparently
accelerate through the profile on $\sim$3.9-day time scales, even
though their recurrence time is shorter.
The saturated ($k_{L}=100$) line exhibits blue-edge variability
on the same time scale.
Figures~11a and 11b also contain the
time-averaged line profiles for Model~1 and the
minimum and maximum absorption templates, which show the extent
of the absorption variability at each velocity.
We also plot the standard deviation of the data, allowing
a qualitative comparison to observed temporal variance spectra.

Figure~12 shows the absorption-column line profile variability for
Model~2.
The accelerating features in this unsaturated ($k_{L}=1$) line
differ from those in Figure~11a in two apparent ways.
First, because of the high-velocity stream induced by the dark spot,
the variability extends out to nearly 5000 km~s$^{-1}$, almost twice
the unperturbed wind terminal speed.
Second, the enhanced absorption at lower line velocities
($v \lesssim 2000 \, \mbox{km s}^{-1}$) represents the mean state,
and the isolated accelerating features appear as a lack of absorption.
Although we anticipated this trend in \S~3.2, it is surprising to
note the overall {\em similarity} between the bright and dark spot
profiles at higher line velocities
($2000 \, \mbox{km s}^{-1} \lesssim v \lesssim 2400 \, \mbox{km s}^{-1}$).
The primary difference here is that the strongest absorption feature trails,
in Model~1, and leads, in Model~2, the overall region of enhanced
absorption.
Far from the star, the near-terminal-speed wind is beginning to
laterally homogenize the bright or dark spot perturbations into a
simpler pattern of alternating large-scale compressions and rarefactions.

In order to interpret more clearly the evolution of DACs in our models,
we utilize the observational fitting technique of Henrichs et
al.\  \markcite{He83}(1983) and Kaper \markcite{K93}(1993),
which represents the DACs as
dense, plane-parallel slabs of gas in the observer's line of sight.
For each component, the dependence of the quotient flux (normalized
by the minimum absorption template) on the
line-of-sight velocity $v$ is fit by
\begin{equation}
  I(v) \, = \, \exp \left\{ - \tau_{c} \exp \left[ - \left(
  \frac{v - v_{c}}{v_t} \right)^{2} \right] \right\} \,\,\, ,
\end{equation}
where $\tau_{c}$ is a representative central optical depth,
$v_{c}$ is the line velocity of the center of the DAC, and $v_{t}$
is related to its width in velocity space.
These three parameters are varied and fit to each feature in the
time series using Marquardt's $\chi^2$ method
(Bevington \markcite{B69}1969).
Although many of our synthetic DACs are asymmetric about $v_{c}$,
with slightly more absorption on the low-velocity side of the feature,
the fits always reproduce well the overall line shape.
One additional useful quantity, the column density of the DAC, is
given for our model lines by
\begin{equation}
  N_{col} \, = \, \frac{\sqrt{\pi}}
  {\kappa_{L} v_{\mbox{\footnotesize th}} \langle m \rangle}
  \frac{\tau_{c} v_{t}}{(1 + v_{c}/c)} \,\,\, ,
\end{equation}
where $\langle m \rangle$ is the mean mass of gas atoms and ions
(see POF\markcite{POF} and Kaper \markcite{K93}1993).

In Figure~13 we plot the resulting fit parameters
($v_{c}$, $v_{t}$, $N_{col}$) as a function of
time for the bright-spot, unsaturated-line models in the parameter study.
The inherent ``overlap'' in the time series (i.e., multiple DACs at
a given time) has been removed to more clearly show the evolution
of the individual DAC.
As is often seen in observed line profile variability, the feature
accelerates through the line profile while growing progressively
narrower, its column density increasing to a maximum value, then
decreasing as the DAC nears its terminal velocity.
In our models, $v_{c}$ often reaches or exceeds the wind's unperturbed
$v_{\infty}$, but the rapidly dropping values of
$N_{col}$ (which also is related to the equivalent
width of the DAC) might preclude actual observation of this final period
of evolution.
In fact, most of the DACs we track seem to approach an initial
``pseudo terminal speed" ($\sim$~0.8-0.9~$v_{\infty}$) while the column
density is at its peak, then accelerate further to the wind's terminal
speed as the column density decreases.

As expected, the DACs produced by radiative-acoustic kinks accelerate
quite slowly.
Figure~13c compares the acceleration of DACs from Models 1, 5A, and
5B with several analytic ``beta'' velocity laws.
We compute $v(t)$ by numerically integrating the kinematic relation
\begin{equation}
  t \, = \, t (r [v]) \, = \,
  \int_{R_{\ast}}^{r} \frac{dr'}{v_{0} + (v_{\infty}-v_{0})
  (1 - R_{\ast}/r')^{\beta}} \,\,\, ,
\end{equation}
where we take $v_{0} = a$ and $v_{\infty} = 2580 \, \mbox{km s}^{-1}$.
The slow acceleration of the DACs is equivalent to
$\beta \approx 2-4$, which agrees with the observations of
Prinja et al.\  \markcite{Pe92}(1992) and
Prinja \markcite{P94}(1994).
Note, however, that an estimation of a single characteristic $\beta$
for a DAC is problematic, since
(1) its terminal speed is not clearly defined, and
(2) $v_{c}(t)$ experiences several minor acceleration and deceleration
episodes superimposed on the overall DAC acceleration.
Figure~14 plots acceleration versus velocity for the DAC of Model~1,
and compares it to the beta laws defined above.
Note the similarity in both magnitude and nonmonotonic behavior between
this theoretical acceleration and that found by Prinja
et al.\  \markcite{Pe92}(1992) for DACs in the wind of $\zeta$~Puppis.

\section{Summary, Conclusions, and Future Work}

We have carried out two-dimensional hydrodynamical simulations of an
azimuthally inhomogeneous radiation-driven wind from a rotating O star.
The wind responds to a photospheric radiative force enhancement by
forming large-scale corotating structures extending far beyond the
region of direct perturbation.
Although classical CIR compressions and rarefactions are often seen,
the most important structures observationally are slowly-propagating
radiative-acoustic kinks or plateaus which have a large Sobolev
optical depth.
These plateaus show up as strong DACs when formed behind streams
resulting from {\em enhanced
mass loss,} and they accelerate at a slower rate than the wind
passes through them.
Preliminary P~Cygni line profile synthesis has shown several important
trends in the emergent DACs from these structures:
\begin{enumerate}
\item The slow acceleration of DAC features is fit reasonably well by
$\beta \approx 2-4$, and we find no significant correlation between
their acceleration time scale and the star's rotation velocity.
Of course, since the CIRs in our models are linked to the rotating
surface, there is a definite correlation between the {\em recurrence}
time scale of DACs and $V_{\mbox{\footnotesize eq}}$.
\item Despite minor variations, the DAC parameters $v_{c}$ and $v_{t}$
do not depend strongly on the amplitude $A$ or full width $\Phi$ of
the model spot perturbation.
The primary exception is the high-amplitude Model~3C ($A=2.50$) which
accelerates much more slowly due to the higher density in the
CIR shock.
\item The optical depth $\tau_{c}$ and column density $N_{col}$
seem to be sensitive probes of the amplitude of the initial
surface perturbation (see Figure~13a, bottom panel).
The concavity of $N_{col}$ during the strongest period of DAC
evolution may be able to provide information about the azimuthal
size of the perturbation, but this is not
as clearly observable an effect
(see Figure~13b, bottom panel).
\end{enumerate}

Let us now compare our model results with actual observations of large-scale
wind structure.
In many cases both (1) slow and quasi-episodic DACs and (2) faster
periodic modulations are seen simultaneously in OB-star winds
(Kaper \markcite{K93}1993; Massa et al.\  \markcite{Me95}1995).
As seen above, the first type of variation can be readily reproduced
by the CIR and radiative-acoustic plateau that results from an
azimuthally localized mass-loss enhancement.
Though our models assume strict rotational periodicity,
such DACs could also be readily caused by {\em transient} CIRs
for which the surface mass loss ``eruption'' lasts long enough
to make structure that covers a substantial portion of the stellar disk.
The second type of structure observed in OB winds (fast,
near-sinusoidal flux variations accelerating {\em with} the wind) is
more difficult to produce with a CIR model,
because the required optical depth or density variations disrupt
the mean wind velocity streaklines.
To explain these faster modulations, we intend to investigate models in
which wind structure is induced by nonradial pulsation (NRP) of
the underlying star
(Owocki, Cranmer, \& Fullerton \markcite{Oe95}1995).

The success here in reproducing realistic DACs suggests that the
CIR model warrants further study and development.
One important extension will be to use a more complete radiative force
that incorporates the line-driven instability.
The resulting stochastic variations may disrupt the large-scale
CIR structure, but slowly evolving kink-like plateaus
have been seen to survive and propagate
in various one-dimensional instability simulations
(Fullerton \& Owocki \markcite{FO92}1992;
Owocki, Fullerton, and Puls \markcite{Oe94}1994).
In addition, extending the present models to three dimensions
may shed light on the variability of wind-compressed disks and other
latitudinally-varying structures in, e.g., Be-star winds.
Finally, it will be important
to develop techniques to synthesize
other observational diagnostics, such as subordinate-level UV
and optical lines, infrared photometry, and continuum and line
polarization, which will allow further constraints
on models of wind variability.
The resulting phenomenological ``atlas'' would provide a solid basis
for interpreting the great diversity of OB-star wind variability
observations in terms of fundamental, dynamical models of
time-dependent wind structure.

\acknowledgments
This work was supported in part by
NSF grant AST 91-15136 and
NASA grant NAGW-2624 (to SPO),
and a NASA Graduate Student Researcher's Program fellowship (to SRC).
Supporting computations were made possible by an allocation from
the San Diego Supercomputer Center.
We thank A.\  Fullerton, K.\  Gayley, D.\  Massa, and D.\  Mullan
for many helpful discussions and comments.
We also thank J.\  Blondin for initially providing the VH-1
hydrodynamics code.

\newpage

\begin{table}
\begin{center}

\begin{tabular}{ccccccccc}
Model & $A$ & $\Phi$ & $V_{\mbox{\footnotesize eq}}$ (km s$^{-1}$) &
max ($\rho / \rho^{(0)}$) & $\Delta v_{r}$ \tablenotemark{a}
(km s$^{-1}$) & $r_{L}/R_{\ast}$ \tablenotemark{b} &
$r_{NL}/R_{\ast}$ \tablenotemark{c} &
$r_{i}/R_{\ast}$ \tablenotemark{d} \\
\tableline
\hspace{0.01in} & \hspace{0.01in} & \hspace{0.01in} &
\hspace{0.01in} & \hspace{0.01in} & \hspace{0.01in} &
\hspace{0.01in} & \hspace{0.01in} & \hspace{0.01in} \\
1  & +0.50 & 20\arcdeg & 230 & 3.747 & -629.2 & 1.607 & 6.417 & 6.736\\
2  & -0.50 & 20\arcdeg & 230 & 5.071 & +2272. & 2.188 &  ---  & 7.646\\
\hspace{0.01in} & \hspace{0.01in} & \hspace{0.01in} &
\hspace{0.01in} & \hspace{0.01in} & \hspace{0.01in} &
\hspace{0.01in} & \hspace{0.01in} & \hspace{0.01in} \\
3A & +0.01 & 20\arcdeg & 230 & 1.023 & -29.30 & 8.169 & 6.005 & 306.0\\
3B & +0.10 & 20\arcdeg & 230 & 1.243 & -193.0 & 2.782 & 6.157 & 31.10\\
3C & +2.50 & 20\arcdeg & 230 & 39.95 & -1285. & 1.210 & 6.157 & 1.981\\
\hspace{0.01in} & \hspace{0.01in} & \hspace{0.01in} &
\hspace{0.01in} & \hspace{0.01in} & \hspace{0.01in} &
\hspace{0.01in} & \hspace{0.01in} & \hspace{0.01in} \\
4A & +0.50 & 10\arcdeg & 230 & 2.840 & -598.3 & 1.163 & 3.736 & 3.868\\
4B & +0.50 & 40\arcdeg & 230 & 2.206 & -493.3 & 2.526 & 10.38 & 12.47\\
4C & +0.50 & 80\arcdeg & 230 & 1.977 & -238.8 & 4.383 & $>$30 & 23.94\\
\hspace{0.01in} & \hspace{0.01in} & \hspace{0.01in} &
\hspace{0.01in} & \hspace{0.01in} & \hspace{0.01in} &
\hspace{0.01in} & \hspace{0.01in} & \hspace{0.01in} \\
5A & +0.50 & 20\arcdeg & 130 & 3.969 & -783.4 & 1.892 & 9.324 & 11.90\\
5B & +0.50 & 20\arcdeg & 330 & 2.485 & -482.9 & 1.492 & 4.934 & 4.556
\end{tabular}
\end{center}

\tablenotetext{a}{We define $\Delta v_{r}$ as the maximum absolute
perturbation ($v_{r} - v_{r}^{(0)}$), either positive or negative.}
\tablenotetext{b}{The CIR compression (feature III) and the Abbott
kink (feature V) branch out together at the linear bifurcation
radius $r_{L}$.}
\tablenotetext{c}{The prograde dense precursor (feature II) and the
CIR rarefaction (feature IV) branch out together at the nonlinear
bifurcation radius $r_{NL}$.}
\tablenotetext{d}{See equation~(\ref{eq:ri}).}

\caption{Summary of Model Parameter Study}
\end{table}

\newpage
\noindent

\notetoeditor{I would have liked to use the figcaption environment for
our figure captions, but I need to specify separate captions for
sub-figures (separate postscript files) under a given figure number.
The figcaption environment unfortunately updates the numbering counter
every time, and doesn't allow ``a,'' ``b,'' etc.   I will provide
{\em commented} figcaption lines to show what the postscript files
should be.}

{Fig. 1.---
Contours of the star-spot force enhancement for a spot with
full width at half maximum $\Phi= 20\arcdeg$.
The contour levels shown range from 0.1$A$ to 0.9$A$ in intervals
of 0.1$A$.
Overplotted are streaklines (obtained by integrating eq.~[15])
of unperturbed wind models from stars rotating
at 0, 130, 230, and 330 km~s$^{-1}$.
}

{Fig. 2.---
Variation with spot amplitude (for $\Phi= 20\arcdeg$) of
(a) radial velocity and (b) density at $r = 10 R_{\ast}$, for
non-rotating winds.
Solid lines show one-dimensional mCAK solutions,
filled circles show two-dimensional hydrodynamical solutions, and
dashed lines show a simple one-parameter fit to both sets of data.
}

{Fig. 3.---
CIR structure for Model~1, settled to a steady state.
Shown are the (a) density, (b) radial velocity, (c) azimuthal
velocity, and (d) radial Sobolev optical depth, all normalized
to the unperturbed wind initial condition.
}

{Fig. 4.---
Streamlines (left) and streaklines (right) for Model~1,
integrated from 72 equally-spaced points on the star, separated
by $5\arcdeg \,$ intervals.
Dashed lines indicate initial points situated at the center of
the spot enhancement.
}

{Fig. 5.---
Normalized density gray-scale for Model~1, as in Figure~3a.
Overplotted are dashed lines which trace the:  (I) direct spot
enhancement, (II) prograde precursor, (III) CIR compression,
(IV) CIR rarefaction, and (V) radiative-acoustic Abbott kink.
}

{Fig. 6.---
Line plots for Model~1 of the radial variation of (a) radial velocity
and (b) density in the equatorial plane at 16 equally-spaced
azimuthal angles, 11$\fdg$25 apart.
}

{Fig. 7.---
As in Figure~3, except for Model~2.
}

{Fig. 8.---
As in Figure~6, except for Model~2.
}

{Fig. 9a.---
Velocity minima and maxima corresponding to CIR compression and
radiative-acoustic kink features, shown for Model~1 and
variable amplitude  Models 3A, 3B, 3C.
The heavy dashed line is the streakline for the unperturbed
one-dimensional standard model wind.
}

{Fig. 9b.---
As in Figure~9a, except for Model~1 and
variable spot-width Models 4A, 4B, 4C.
}

{Fig. 9c.---
As in Figure~9a, except for Model~1 and
variable rotation   Models 5A, 5B.
}

{Fig. 10.---
Radial velocity showing the the radiative-acoustic kink and CIR shock
for models with varying spot amplitude.
Each $\phi$ slice was chosen independently to align the initial
kink deceleration at the same radius.
}

{Fig. 11a.---
SEI absorption-column line variability for Model~1, computed for
an unsaturated line, $k_{L}=1$.
The gray-scale values range from white (maximum flux) to black
(minimum flux), measured relative to the dashed-line templates in
the middle panel.
The solid lines show the time-averaged line profile, in the middle
panel, and the standard deviation, or square root of the variance,
in the bottom panel.
}

{Fig. 11b.---
As in Figure~11a, except for
a saturated line, $k_{L}=100$.
}

{Fig. 12.---
As in Figure~11a, except for Model~2.
}

{Fig. 13a.---
Best-fit DAC features $v_{c}$ (central line velocity),
$v_{t}$ (characteristic width), and
$N_{col}$ (slab-model column depth), shown for
Model~1 and
variable amplitude  Models 3A, 3B, 3C.
The line styles for each set of models are identical to those in
Figure~9.
}

{Fig. 13b.---
As in Figure~13a, except for Model~1 and
variable spot-width Models 4A, 4B, 4C.
}

{Fig. 13c.---
As in Figure~13a, except for Model~1 and
variable rotation   Models 5A, 5B.
}

{Fig. 14.---
DAC acceleration versus velocity for Model~1 (dashed line) and analytic
``beta'' velocity laws (solid lines).
}

\end{document}